\documentclass[IOP]{emulateapj} 
\usepackage{amssymb}
\usepackage{longtable}

 \shorttitle{The Redshift-Dependent Ly$\alpha$
  fraction in LBG samples}
\begin{document}

\shortauthors{Schenker et al.}
\title{Keck Spectroscopy of Faint $3<z<8$ Lyman Break Galaxies:- 
Evidence for a Declining Fraction of Emission Line Sources In the Redshift Range $6<z<8$}


\author {Matthew A. Schenker\altaffilmark{1},  Daniel P. Stark\altaffilmark{2}, Richard
  S. Ellis\altaffilmark{1}, Brant E. Robertson\altaffilmark{1}, James S. Dunlop\altaffilmark{3},
  Ross J. McLure\altaffilmark{3}, Jean-Paul Kneib\altaffilmark{4} \& Johan Richard\altaffilmark{5}
  }

 \altaffiltext{1}{Department of Astrophysics, California
  Institute of Technology, MS 105-24, Pasadena, CA 91125}
\altaffiltext{2}{Kavli Institute of Cosmology \& Institute of
  Astronomy, University of Cambridge, Madingley  Road, Cambridge CB3
  0HA, UK}  
  \altaffiltext{3}{Institute for Astronomy, University of Edinburgh, Royal Observatory, 
Edinburgh EH9 3HJ, UK}
  \altaffiltext{4}{Laboratoire d'Astrophysique de Marseille, F-13388 Marseille cedex 13, France}
  \altaffiltext{5}{Observatoire de Lyon, F-69561, Saint-Genis-Laval, France  }

\begin{abstract} 
 
Using deep Keck spectroscopy of Lyman break galaxies
selected from infrared imaging data taken with the Wide Field Camera 3
onboard the Hubble Space Telescope,  we present new evidence for a 
reversal in the redshift-dependent fraction of star forming galaxies with 
detectable Lyman alpha (Ly$\alpha$) emission in the redshift range $6.3 < z <8.8$.  Our earlier surveys with the
DEIMOS spectrograph demonstrated a significant increase with redshift 
in the fraction of line emitting galaxies over the interval $4<z<6$, particularly 
for intrinsically faint systems which dominate the luminosity density. Using 
the longer wavelength sensitivities of LRIS and NIRSPEC, we have targeted 19
Lyman break galaxies selected using recent WFC3/IR data whose photometric redshifts are in 
the range $6.3<z<8.8$ and which span a wide range of intrinsic 
luminosities. Our spectroscopic exposures typically reach a 5$\sigma$ 
sensitivity of $<50$ \AA\ for the rest-frame equivalent width (EW) of Ly$\alpha$ 
emission. Despite the high fraction of emitters seen only a few hundred million years
later, we find only 2 convincing and 1 possible line emitter in our more distant sample. 
Combining with published data on a further 7 sources obtained using FORS2 on the ESO VLT,
and assuming continuity in the trends found at lower redshift, 
we discuss the significance of this apparent reversal in the redshift-dependent 
Ly$\alpha$ fraction in the context of our range in continuum luminosity. Assuming
all the targeted sources are at their photometric redshift and our assumptions about the Ly$\alpha$
EW distribution are correct, we would expect
to find so few emitters in less than 1\% of the realizations drawn from our lower
redshift samples. Our new results provide 
further support for the suggestion that, at the redshifts now being probed spectroscopically, 
we are entering the era where the intergalactic medium is partially neutral.
With the arrival of more sensitive multi-slit infrared spectrographs, the prospects 
for improving the statistical validity of this result are promising.

\end{abstract} 

\section{Introduction}
\label{sec:intro}

Determining when neutral hydrogen in the intergalactic medium (IGM)
was reionized is an important question in observational cosmology and
a precursor to understanding whether star forming galaxies provided the
necessary UV photons \citep{Robertson10}. One of  the most practical probes of reionization 
with current facilities utilizes the frequency of occurrence of Ly$\alpha$ emission in 
star forming galaxies.  As Ly$\alpha$ photons are resonantly scattered 
by neutral hydrogen, the abundance of Ly$\alpha$ emitters should decrease 
as observations probe into the era where there is neutral gas (e.g., \citealt{Malhotra04}, 
\citealt{Ouchi10}, \citealt{Kashikawa11}).  The
recent discovery of large numbers of candidate galaxies beyond
$z\simeq7$ through multi-color imaging undertaken with the infrared 
Wide Field Camera (WFC3/IR) onboard Hubble Space Telescope (HST) 
(e.g., \citealt{Bouwens10b,Bunker10,Finkelstein10,McLure10}) now makes it feasible to track the occurrence of 
Ly$\alpha$ line emission to interesting redshifts where neutral hydrogen may be present.

Of course, astrophysical factors other than a neutral IGM can also 
affect the presence of Ly$\alpha$ emission. Because of this, an alternative approach
for gauging when reionization occurred, introduced by \citet{Stark10} (hereafter Paper I),
is to spectroscopically measure {\it the fraction of Ly$\alpha$ emission within color-selected 
Lyman Break Galaxy (LBG) populations}.  Tracking the redshift-dependent fraction in
a well-defined population avoids consideration of absolute changes in the host galaxy 
number density such as has been the case in studies based on the luminosity function of 
narrow-band selected Lyman $\alpha$ emitters (LAEs). Furthermore, 
evolution in dust obscuration can be independently tracked via correlations seen with 
the colors of the rest-frame UV continuum. Paper I presented a comprehensive 
survey of over 600 LBGs with deep spectra, mostly 
undertaken with the DEIMOS instrument on Keck, but including published samples from 
the Very Large Telescope (VLT, \citet{Vanzella09} and references therein).
In that paper we demonstrated the utility of the method
and discussed the paucity of line emission in gravitationally-lensed
$z>7$ candidates from the sample of \citet{Richard08}. 

In \citet{Stark11}  (hereafter Paper II), through ultra-deep exposures with DEIMOS
we significantly improved the line emission statistics at $z\simeq 6$, providing  
a robust measure of the rest-frame EW distribution of
Ly$\alpha$ emission at the highest redshift when the Universe is believed to be fully ionized (\citealt{Fan06}, c.f. \citealt{Mesinger10}). 
This provides a sound basis for predicting the likelihood of emission at higher redshift
and thereby enabling a test of whether there is absorption by neutral gas. Significantly, 
we found that over 50\% of moderately-faint ($-20.25< \rm{M}_{UV} <-18.75$) $z\simeq 6$
LBGs exhibit strong emission with rest frame EWs $> 25$ \AA. As this
fraction increases over $4<z<6$, we argued on continuity grounds that
we should expect a high success rate in recovering line emission from the newly-found
WFC3/IR samples of $z>7$ LBGs {\it unless} we encounter a more neutral IGM in the short
time interval prior to $z\simeq 7$.  Some evidence for this is seen in the recent studies of \cite{Fontana10}
and \cite{Vanzella11}.

The present paper is concerned with an initial application of this test
to the newly-available sample of WFC3/IR candidates with photometric redshifts 
in the redshift range $6.3<z<8.8$. Two important factors have motivated and shaped our
program. Firstly, it is important to note that Ly$\alpha$ emission is the only
spectroscopic redshift indicator for galaxies beyond $z\simeq 6$.  Since it is the absence
of strong Ly$\alpha$ emission that provides the basis for considering an increased 
neutral fraction, it is important to be sure that the targets are truly at the expected redshifts.
Many early candidate LBGs believed to lie beyond $z\simeq 6-7$ remained controversial 
because of their limited or marginal photometry.  The improved filter set and superior performance 
of WFC3/IR has given us confidence that the current list of $z\simeq 7-8$ candidates is more 
robust than those based on earlier NICMOS data \citep{Robertson10}.  Secondly, to match the lower
redshift data, sampling from a similarly wide range of LBG luminosities, as we do here, will be advantageous.  As shown in 
Papers I and II, the fraction of line emission increases in intrinsically fainter 
sources and so by comparing fractions with respect to their LBG luminosities,
we may gain additional evidence for the onset of the neutral era.

Throughout the paper, we adopt a $\Lambda$-dominated, flat universe
with $\Omega_{\Lambda}=0.7$, $\Omega_{M}=0.3$ and
$\rm{H_{0}}=70\,\rm{h_{70}}~{\rm km\,s}^{-1}\,{\rm Mpc}^{-1}$. All
magnitudes in this paper are quoted in the AB system \citep{Oke83}.

\section{Observations}

In compiling a target list for this program, we are guided by the need
for a robust photometric redshift for each galaxy based on improved
photometry from WFC3/IR and a range of rest-frame
UV luminosities ($\rm{M}_{UV}$\footnote{corresponding to a rest wavelength
$\lambda\simeq 1500$ \AA\ (Paper I)}). Our primary source of targets 
for the wide-field multi-slit capabilities of the Low Resolution Imaging Spectrometer 
(LRIS) on the Keck I telescope \citep{Oke95} equipped with a new red-sensitive
CCD was $i'$ and $z'$-drop candidates whose photometric redshifts $z> 6.3$
from the HST Early Release Science (ERS) field \citep{Hathi10, McLure11}.  The grating
for these observations was blazed at 600 lines mm$^{-1}$.  On January 7 and 
February 4 2011 we secured 7 hours of on-source integration for 8
suitable targets on a single mask using slit widths of 1$^{\prime\prime}$, 
observed through a median seeing of 0.98$^{\prime\prime}$.

In a more ambitious campaign probing to higher redshift we also targeted 
3 $z'$-drop sources from the Hubble Ultradeep Field (HUDF) P34 field (GO 11563, 
PI: Illingworth) and an additional gravitationally-lensed source in the cluster MS0451-03
(GO 11591, PI: Kneib) using the near-infrared spectrograph NIRSPEC \citep{McLean98} during
November 14-17 2010 and Jan 14-15 2011. This extends our search for Ly$\alpha$ emission
up to a redshift $z\simeq 8.2$. Although we undertook extended
integrations on all 4 sources with a 0.76$^{\prime\prime}$ slit, tracking difficulties
affected some exposures. To determine the effective on-source integration time, 
we secured our astrometric position for each exposure by locating objects visible in 
the slit viewing camera to a precision of $\sim$ 0.2$^{\prime\prime}$.  
We continued this campaign over May 15-18 2011.  During these 4 nights,
we did not encounter any tracking difficulties and, in excellent conditions, successfully 
used NIRSPEC to study an additional 7 WFC3-IR dropouts drawn from numerous surveys: the BoRG pure parallel
survey \citep{Trenti11} (also independently discovered by \citealt{Yan11}), the EGS region in CANDELS (\citealt{Grogin11}; \citealt{Koekemoer11}), and the 
lensing clusters Abell 1703 (GO 10325, PI: Ford, \citealt{Bradley11}) and Abell 2261 (CLASH survey, \citealt{Postman11}). 

We reduce the LRIS data following standard procedures, with bias subtraction and 
flat-fielding using dome exposures. We used the \cite{Kelson03} code to remove 
spatial and spectral distortion and to model and subtract the sky emission. Wavelength 
calibration was determined directly from sky lines. A final two-dimensional spectrum was 
extracted for each object with pixels binned logarithmically by 
$\Delta$log$(\lambda) = 4.02 \times 10^{-5}$.  As in Papers I and II, we search through 
the two-dimensional spectrum visually to identify emission lines, and confirm these with
a boxcar extracted one-dimensional spectrum.

Our exposures with NIRSPEC were conducted with typical 
spatial dithering of 5$^{\prime\prime}$.  In the case of some lensed sources, we
dithered by longer amounts to ensure the arc was oriented along the slit, and was
not spatially coincident with any background objects.  We flat-fielded and 
sky-subtracted the spectra using IDL routines written by G. Becker (2010, private 
communication).  To compute the camera distortion and spectral curvature, we fit 
traces of standard stars along the slit, and skylines perpendicular to the slit.  From this,
we then derive a wavelength and sky position for all pixels in each two-dimensional 
spectrum, which were used to align the individual exposures.  A final spectrum was 
created for each object by median stacking all exposures to eliminate signals from 
cosmic ray strikes.

\medskip
\begin{table*}

\caption{Catalog of sources with Keck spectroscopy and emission line properties, when detected.  1$\sigma$ 
magnitude errors are listed in parenthesis.  For non-detections, we list 2$\sigma$ limiting magnitudes.
For our NIRSPEC targets, times correspond to exposures in filters N1 and N2, which cover wavelength 
ranges of 9470 - 11210 \AA\ and 10890 - 12930 \AA, respectively.}

\medskip
\small
\begin{tabular}{l c c c c c c c c c c}
\hline
\hline
ID & R.A. & Dec & $z_{850}$  & $J_{125}$ & $H_{160}$ & $\mu$\footnotemark[1] & $z_{phot}$ & t$_{exp}$ [hr] & $z$ & EW [\AA] \\
\hline
 & & & & LRIS \\
\hline 
ERS 5847\footnotemark[4]\footnotemark[8] & 03:32:16.0 & -27:43:01.4  &$ 26.6(0.1) $\footnotemark[2] &  $ 26.6(0.1) $ &  $ 26.7(0.1) $ & -& $6.48$  & $7$ &  \\
ERS 7376\footnotemark[4]\footnotemark[8] & 03:32:29.5 & -27:42:04.5  & $ 27.2(0.1) $\footnotemark[2] &  $ 27.0(0.1) $ &  $ 27.0(0.1) $  &-& $6.79$  & $7$ &\\
ERS 7412 & 03:32:10.0 & -27:43:24.0   & $ 26.9(0.1) $\footnotemark[2] &  $ 27.0(0.1) $ &  $ 26.7(0.1) $ &-&  $6.38$ & $7$ &\\
ERS 8119 & 03:32:29.5 & -27:41:32.7 & $ 27.7(0.2) $\footnotemark[2] &  $ 27.1(0.1) $ &  $ 27.5(0.1) $ & -& $6.78$ & $7$ &\\
ERS 8290 & 03:32:13.4 & -27:42:30.9  & $ 27.3(0.1) $\footnotemark[2] &  $ 27.1(0.1) $ &  $ 26.8(0.1) $ &-& $6.52$ &  $7$ & \\
ERS 8496\footnotemark[4] & 03:32:29.7 & -27:40:49.9   & $ 27.2(0.1) $\footnotemark[2] &  $ 27.3(0.1) $ &  $ 27.5(0.1) $ &-&  $6.52$ & $7$ & $6.441$ & $69 \pm 10 $\\
ERS 10270 & 03:32:29.5 & -27:42:54.0  & $ 28.1(0.1) $\footnotemark[2] &  $ 27.4(0.1) $ &  $ 28.0(0.2) $  & -& $7.02$ & $7$ &\\
ERS 10373 & 03:32:27.0 & -27:41:42.9  & $ 27.5(0.1) $\footnotemark[2] &  $ 27.4 (0.1) $ &  $ 27.8(0.2) $  & -& $6.44$ & $7$ &\\
\hline
 & & & & NIRSPEC\\
 \hline
A1703\_zD1\footnotemark[5]  & 13:14:59.4 & 51:50:00.8 &  $25.8(0.2)$  & $24.1(0.1)$ & $24.0(0.1)$ & $9.0$   &$6.75$ & $2,-$\\
A1703\_zD3\footnotemark[5]  & 13:15:06.5 & 51:49:18.0 &  $26.8(0.5)$  & $25.5(0.1)$ & $25.1 (0.2)$ & $7.3$   & $6.89$ & $2,-$\\
A1703\_zD6\footnotemark[5]  & 13:15:01.0 & 51:50:04.3 &  $27.9(0.5)$  & $25.8(0.1)$ & $25.9(0.1)$ & $5.2$   & $7.02$ & $5$,  $3$ & $7.045$ & $65 \pm 12$\\
A1703\_zD7\footnotemark[5]  & 13:15:01.3 & 51:50:06.1 &  $> 28.5$      & $26.8 (0.2)$ & $26.4 (0.2)$ & $5.0$   & $8.80$ & $5$,  $3$\\
A2261\_1    & 17:22:28.7 & 32:08:30.9 &  $>28.6$  & $26.9(0.1)$& $27.3 (0.1)$ & $3.5$    & $7.81$ &$5.7,-$ & \\
BoRG\_58\_1787\_1420\footnotemark[6]     & 14:36:50.6 & 50:43:33.6  & $>27.9$\footnotemark[2] & $25.8(0.1)$ & $25.9 (0.2)$ & -   & $8.27$& $2$, $3$ &\\
EGS\_K1     & 14:19:24.2 & 52:46:36.2 & $>27.8$\footnotemark[3] &  $25.3(0.1)$ & $25.4 (0.1)$ & - &  $8.27$ & $2.5,-$ & \\
HUDF09\_799\footnotemark[7] & 03:33:09.1 & -27:51:55.4  & $>29.1$ &  $27.7(0.1)$ & $ 27.6 (0.2)$ & - & $6.88$  & $4.5,-$\\
HUDF09\_1584\footnotemark[7] & 03:33:03.8 & -27:51:20.4 & $27.2(0.1)$ & $26.7 (0.1)$ & $26.6 (0.1)$ & - &  $7.17$ & $5.5,-$\\
HUDF09\_1596 & 03:33:03.8 & -27:51:19.6 & $27.3(0.1)$ &  $26.8 (0.1)$ & $26.8 (0.1)$ & - &  $7.45$ & $5.5,-$ & $6.905?$ & $30  \pm 15 $\\
MS0451-03\_10 & 04:54:08.8 & -3:00:29.1 & $>28.3$ & $26.7(0.1)$ & $26.9 (0.1)$ & $50$ &  $7.50$ & $2.5,-$\\  
\hline
\hline
\end{tabular}

\footnotetext[1]{Best fit magnification for sources in our sample which are gravitationally lensed.}
\footnotetext[2]{Y$_{098M}$}
\footnotetext[3]{I$_{814W}$}
\footnotetext[4]{In order listed in table, discovered as Objects 1,4,3 in \cite{Wilkins10}}
\footnotetext[5]{Discovered in \cite{Bradley11}}
\footnotetext[6]{Discovered in \cite{Trenti11}}
\footnotetext[7]{In order listed in table, discovered as P34.z.6106, P34.z.4809 in \cite{Wilkins11}}
\footnotetext[8]{In order listed in table, identified as ERS.z.45856 and ERS.z.87209 in \cite{Wilkins11}}

\label{params}
\end{table*}

In total, this paper therefore presents the results of Keck spectroscopy for 19 WFC3-IR selected 
sources whose photometric redshifts lie in the range $6.3<z<8.8$. A summary of the new observations 
is given in Table 1. To this sample, we add a further 7 $z>6.3$ sources discussed by \citet{Fontana10}.
Figure 1 compares the UV absolute magnitude distribution of the combined sample with that presented 
for the redshift range $z\simeq 5-6$ in Paper II; clearly the samples span a similar luminosity range.  This
luminosity range is broader than the recent work of \cite{Ono11} and \cite{Pentericci11}.  In similar 
spectroscopic follow-up campaigns, they target brighter dropouts, primarily with M$_{UV} < -21.4$, and 
$-21.75 < $M$_{UV} < -20.0$, respectively.

\begin{figure}
\begin{center}
\includegraphics[width=0.5\textwidth]{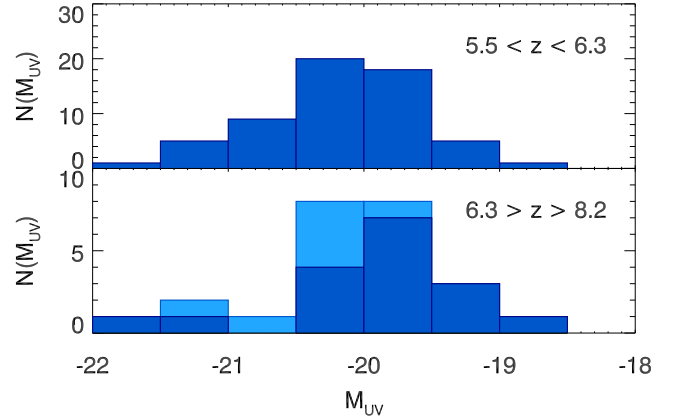}
\caption{Distribution of rest-frame UV absolute magnitudes, $M_{UV}$, for $i'$-drop sources discussed in 
Paper II with z $\sim$ 5.5-6.3 (top panel) compared with those for the present survey of $i'$ and $z'$-drops 
at z $\geq$ 6.3 (lower panel). There is an additional dropout, A2261\_1, not shown on this histogram at 
M$_{UV} \simeq -16$.  Dark shading in the lower panel refers to sources selected on the
basis of WFC3/IR imaging in the Keck campaign, (Table 1); light shading 
refers to additional data drawn from the VLT campaign of \citet{Fontana10}.}
\label{fig:target_hist}
\end{center}
\end{figure}

Remarkably, from the new Keck exposures, we find very few convincing detections of line emission.
Figure 2 (bottom panel) shows a 2-D spectroscopic montage of 4 sources for which line emission may be present, of which
one case (HUDF09\_1596) is marginal (2$\sigma$) and the other (ERS 8290) lies outside the expected 
redshift range if it is Ly$\alpha$. 

The emission line apparently seen in ERS 8290 (a $z'$-band dropout) is detected at $>$5$\sigma$ with a flux of 
6.7 $\pm$ 0.8  $\times$ $10^{-18}$ erg cm$^{-2}$ s$^{-1}$ at $\lambda$7644 $\pm$ 2 \AA. 
It also exhibits the asymmetric profile characteristic of Ly$\alpha$ in the 1D extraction, but this would
place the object at $z=5.286$, quite discrepant from our photometric estimate of $z$ = 6.52.  
However, upon examining the positioning of our LRIS slits more carefully (bottom panel of Figure 2), 
we find there is a faint $V$-drop candidate with $i'$ = 27.5 and $z'$ = 27.9 only $~0.4^{\prime\prime}$ 
away which would have been at least partially visible through our slit during the exposure. After 
subtracting the line flux from the $i'$-band photometry, we determine a photometric redshift 
of $z$ = 4.91, in reasonable agreement with the Keck spectroscopy, particularly given a greater 
line flux (as is likely given the object's poor positioning in the slit) would increase the 
photometric redshift estimate. The resolution of the confusion arising from these two proximate sources
emphasizes again the prominence of Ly$\alpha$ emission in low luminosity $z\simeq 5-6$
sources (Papers I and II).

The two satisfactory detections refer to emission at $\lambda$9045 $\pm 2$ \AA\ for ERS 8496 in the LRIS mask 
and emission at $\lambda$9780 $\pm\ 4$ \AA\ in the NIRSPEC exposure of the lensed source Abell 1703\_zD6.  
Both objects are detected at $\geq$ 5$\sigma$ in our final exposures.  In our 1D extraction
of ERS 8496, the emission line has a FWHM of 9 $\pm$ 1 \AA, and displays an asymmetric profile with a steeply rising blue edge and slowly
decaying red tail, characteristic features of Ly$\alpha$ at high redshift.  Because our spectral resolution is 
significantly lower (6.5 \AA\ for NIRSPEC versus 4.1 \AA\ for LRIS) in our spectrum of A1703\_zD6, so we are
unable to determine any line profile information.  The emission feature is seen independently in
coadditions on two separate nights, indicating its reality.

If both lines are Ly$\alpha$, the implied spectroscopic redshifts for ERS 8496 and 
Abell 1703\_zD6  are $z=6.441 \pm 0.002$ and $7.045 \pm 0.004$, respectively, in excellent agreement with our photometric 
predictions of 6.52 and that of 7.0 derived by \cite{Bradley11}.  The measured line fluxes for the two objects
are 9.1 $\pm$ 1.4 and 28.4 $\pm$ 5.3 $\times$ $10^{-18}$ erg cm$^{-2}$ s$^{-1}$.   We then assume a
spectral slope of $\beta$ = -2, which is characteristic of galaxies at this redshift \citep{Dunlop11}, though 
there may be evidence for steeper slopes at sub-$L_{\star}$ luminosities \citep{Bouwens10a}.  Under this 
assumption, taking the magnitude from the first filter in which the object is detected ($Y_{098M}$ for ERS 8496, 
and $J_{125}$ for 1703\_zD6), we calculate EWs of 69 $\pm$ 10 and 65 $\pm$ 12 \AA, respectively.  
Because our objects have additional coverage longward of the detection filter, we can also compute a value for $\beta$,
and extrapolate to find the continuum flux at $\lambda_{rest} = 1216$ \AA.  Using the formulae of \cite{Dunlop11},
we find $\beta$ = -2.39 $\pm$ 0.55, and -2.44 $\pm$ 0.64.  When computing EWs using this method, we obtain
67 $\pm$ 11 and 59 $\pm$ 12 \AA, respectively.

\begin{figure}
\begin{center}
\includegraphics[width=0.5\textwidth]{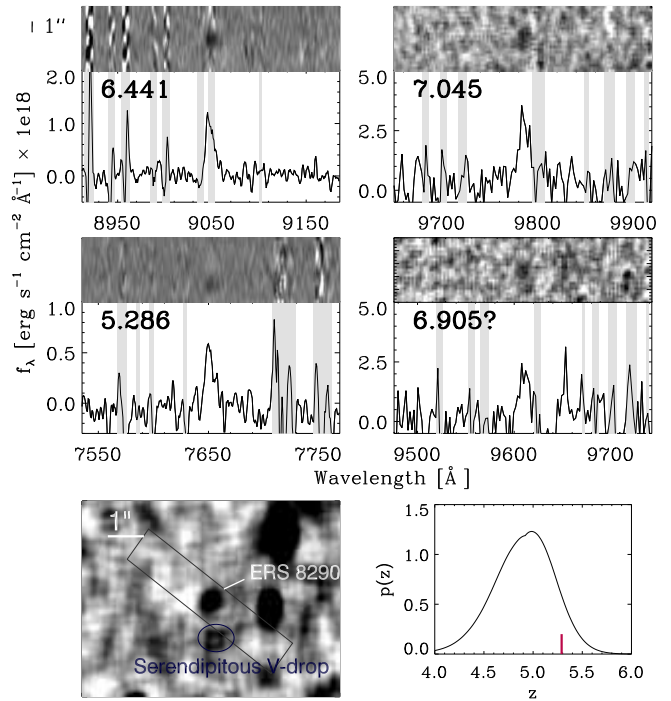}
\vspace{0.1in}
\caption{Montage of Ly$\alpha$ emission detected from 4 sources in our
Keck survey, along with boxcar-extracted 1D spectrum.  Wavelength ranges
contaminated by strong skylines are shaded in grey in the 1D extraction.
The top row shows the two robust detections of ERS 8496 and A1703\_zD6 at $z=6.441$ 
and $z$=7.045 respectively. The bottom row shows a marginal detection at $z$=6.905
for HUDF09\_1596 and a likely $Ly\alpha$ line at $z=5.286$ arising from a serendipitous 
$V$-drop close to ERS 8290 as illustrated in the bottom left slit image.  The bottom right panel
shows the photometric redshift distribution for this serendipitous $V$-drop, with a vertical
line indicating the observed spectroscopic redshift
(see text for further details).}
\label{fig:spec_montage}
\end{center}
\end{figure}

\section{Analysis}

Our approach in this paper is to compare the rate of occurrence of Ly$\alpha$ in our
new $6.3<z<8.8$ sample with that expected from our reference sample of 
$i'$-drops with $5.5<z<6.3$ drawn from Paper II (Figure~1).  In both Papers I and II we showed that the 
rest-frame EW distribution is a function of rest-frame UV absolute magnitude, 
$\rm{M}_{UV}$, and thus we additionally take this luminosity dependence into account. We estimate
the luminosities of all our sources in Table~1 from their photometric redshift and incorporate the
lensing magnification $\mu$ for our lensed sources from \cite{Bradley11} for Abell 1703 and from the
mass model of \cite{Richard10} for MS0451-03 and Abell 2261. For our baseline Ly$\alpha$ EW 
distribution, we use the data from of Paper II at $z \simeq 6$, separated into high and low luminosity regimes.

The fraction of emitters within each bin of EW $>$ 25 A is taken directly from Fig. 2 of Paper II.  
We set the slope of the distribution within an EW bin equal to the slope between 
the two lowest bins in Paper II, 25 \AA\ $< $ EW $ <$ 55 \AA , and 55 \AA\ $< $ EW $<$ 85 \AA.  This slope is 
equal to $dp($EW$)/d$EW = $-0.0030$ for the lower luminosity sample ($-20.25 < $M$_{UV} < -18.75$), and $-0.0017$ for 
the higher luminosity sample ($-21.75 < $M$_{UV} < -20.25$).  
To create the probability distribution for galaxies with EW less than 25 \AA, we extrapolate to 
EW = 0 \AA\ using this slope, and assign the remaining fraction of galaxies as non-emitters.
In Papers I and II we also showed the fraction of emitters is a function of redshift, rising significantly for 
lower luminosity sources over $4<z<6$, most likely as a result of reduced dust extinction in the early Lyman break 
population. Therefore, as discussed in Paper II, we have also used a projected rest-frame
EW distribution at $z\simeq 7$ assuming this evolutionary trend continues beyond $z\simeq$ 6.

Two key factors enter into the calculation of the visibility of
line emission in a ground-based survey. Firstly, for any target with a particular photometric redshift
likelihood function $p(z)$, it may be that the spectral region surveyed by LRIS or NIRSPEC
does not completely cover the expected wavelength range where Ly$\alpha$ might be
present. Secondly, the EW limit for Ly$\alpha$ emission will be a highly non-uniform 
function of wavelength due to the mitigating effect of night sky emission. Provided the
photometric redshift solution we derive is robust, we can estimate both factors
and hence derive the likelihood of seeing Ly$\alpha$ for each of our 26 sources assuming
the relevant wavelength range studied and the exposure time secured, if the particular
source of a given $M_{UV}$ has a EW distribution drawn from the sample with $5.5<z<6.3$.

In the case of those sources for which the wavelength range searched does not fully sample 
the extended $p(z)$, we reduce the detection likelihood by the fraction of the integrated
$p(z)$ that lies outside our search range. For each target, we determine its redshift 
probability function $p(z)$ using the photometric redshift code EAZY \citep{Brammer08}.
To determine the varying visibility function within our search range, we first estimate the
the noise within an aperture encompassing the expected profile of the line, assuming  
an emission line width of 10 \AA\ FWHM which is typical of  those detected in Paper II. 
Figure 3 illustrates the 5$\sigma$ EW limit as a function of wavelength (and Ly$\alpha$ redshift) 
for most sources in our 12 hour LRIS exposure ($6.3<z<7.2$) and a typical source studied with 
NIRSPEC ($6.8<z<8.2$) during a 5 hour exposure.  We note that although our survey spans 
a large range in redshift ($6.3<z<8.8$), the $p(z)$ distributions 
for individual galaxies typically span a much smaller range.  The average
1$\sigma$ redshift confidence interval for sources in our sample is only $\Delta z$ = 0.43.

Since the NIRSPEC exposures were
usually single-object exposures, the limits vary from source to source depending on
the conditions and exposure times. We then apply a completeness correction to account for 
the fact that an emission line may fall in a noise trough and lie undetected, despite having an 
intrinsic flux above the 5$\sigma$ limit.  To estimate this completeness correction, we follow 
the methods discussed in detail in Paper II, where
we simulate the addition and recovery  of  fake line emission in our actual spectra, again 
assuming a FWHM of 10 \AA. As the absolute limits vary from source to source our Keck 
survey is not complete to a fixed EW limit but, provided the limits are well-understood for 
each source, we can readily estimate the probability of seeing Ly$\alpha$ in our exposures. 
In the case of the \citet{Fontana10} FORS2 survey we estimated the night sky emission from our own LRIS
exposures normalizing the limits from numerical data supplied in that paper.

\begin{figure}
\begin{center}
\includegraphics[width=0.5\textwidth]{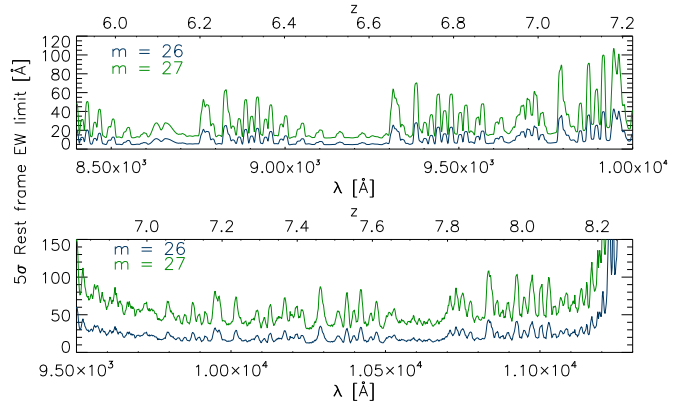}
\vspace{0.1in}
\caption{Sensitivity limits to Ly$\alpha$ emission in our new Keck spectroscopic campaign:
The panels show the 5$\sigma$ limiting EW calculated for a typical source
studied in our 12 hour multi-slit LRIS exposure (top) and an example 5 hour long slit NIRSPEC observation
(bottom). The limits vary from source to source depending on the continuum brightness and
the exposure times.  An additional completeness correction is taken into account by adding
 and attempting to recover fake emission lines with fluxes equal to the 5$\sigma$ flux limit 
 at the wavelength of insertion.  See text and Paper II for more detailed discussion. }
\label{fig:sensitivity_limits}
\end{center}
\end{figure}

The above simulations can be used to verify that our Keck
survey is well-placed to search for Ly$\alpha$ emission.  Out of the combined 26 targets from our survey
and that of \cite{Fontana10}, 24 are covered spectroscopically over more than half the integrated probability of their
photometric redshift distribution, and 17 are covered over 95\% of the range.  Additionally, we are able to determine
the fraction of our spectra occulted by OH sky emission.  For example, for a J=27 galaxy in one of our LRIS exposures,
we are sensitive to lines with EW $\geq$ 30 \AA\ over $70\%$ of our usable spectral range (see Figure 3).  Similarly,
for a 5 hr NIRSPEC exposure of a J$=27$ galaxy, we are sensitive to lines with EW $>$ 55 \AA\ over $49\%$
our spectral range.

The results of the simulations are shown in Figure 4 where, depending on whether we adopt
the EW distribution observed at $5.5<z<6.3$ or that extrapolated to $z\simeq7$ in Paper II
assuming continuity in redshift-dependent increase in line emission seen over $4<z<6$,
we would expect to recover 7-8 emission lines in the combined Keck and VLT surveys.
In contrast, we have only 2 robust detections (both in the Keck sample, Figure 2) and
at most 4 including the marginal candidate discussed by \citet{Fontana10} and HUDF09\_1596
at $z=6.905$ shown in Figure 2. Assuming all the targeted sources are at $z>6.3$, given our 
previously mentioned assumptions, our results reject the input EW distributions at the 99.3$\%$
level of significance (91.4$\%$ if the two marginal detections are included).

\begin{figure}
\begin{center}
\includegraphics[width=0.5\textwidth]{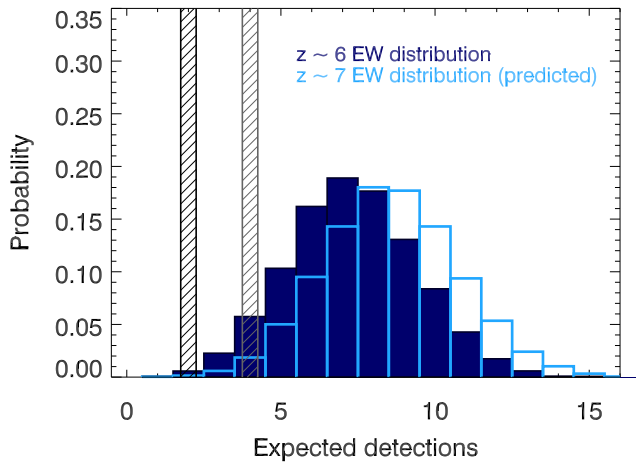}
\vspace{0.1in}
\caption{The expected number of detected Ly$\alpha$ emission lines with greater than
or equal to 5$\sigma$ significance in the combined Keck and VLT survey of 26 sources.
The blue histogram shows the likelihood function for 10,000 Monte Carlo realizations assuming the
intrinsic line emission properties follow the luminosity dependence seen in our $5.5<z<6.3$
$i'$-drop sample (Figure 1 (top). The open histogram shows the expectation if the fraction of
line emitters continues to increase with redshift at the rate described in Paper II. Vertical lines
show the recovered number of emitters (robust and maximal including marginal
detections in both the Keck and VLT surveys).}
\label{fig:monte_carlo_sim}
\end{center}
\end{figure}

We can display the significance of this downturn with increasing redshift in the terms of
the fraction of Ly$\alpha$ emission seen in Lyman break galaxies, $X$(Ly$\alpha$) as in Paper I. 
The difficulty we face in creating such a figure is  the non-uniform EW limit across
the various targets in the Keck and VLT campaigns, in contrast to the more straightforward uniform
search we undertook with DEIMOS at $4<z<6$. To account for this, we assume a simple model in which
Ly$\alpha$ emission is transmitted without IGM absorption for a fraction $f$ of galaxies, while it is 
fully extinguished by the IGM for a fraction of galaxies (1-$f$).  We assume that $f=1$ at redshifts
below 6, where the universe is believed to be highly ionized \citep{Fan06}, and that $f$ is independent of the
intrinsic EW of a Ly$\alpha$ emission line.  We caution that an interpretation in terms of absolute values of $f$
is premature, as there is still some debate on whether the IGM is fully ionized at $z\sim6$ \cite{Mesinger10}, but 
emphasize that our value of $f$ at $z\sim7$ is computed relative to the value assumed at $z\sim6$.  Additionally,
with the increased fraction of emitters in our $z\sim6$ sample from Paper II, we do not see any evidence for a 
decrease in $f$ prior to $z\sim6$, though we cannot rule it out.

It is important to note that our $f$ is differs from $f_{esc}^{Ly\alpha}$, commonly defined in the literature as 
the total escape fraction of Ly$\alpha$ photons (e.g., \citealt{Hayes11}).  $f_{esc}^{Ly\alpha}$ represents
the total transmission of Ly$\alpha$, accounting for both attenuation of photons within the galaxy by mechanisms 
such as dust, as well as any attenuation by the IGM.  Our definition of $f$ is only intended to account for any 
downturn in the fraction of LBGs which show observable Ly$\alpha$ emission from the $z=6$ (or $z=7$) extrapolated
EW distributions from Paper II, and represents an IGM extinction averaged over the entire population.

To compute the most likely value of $f$, we undertake Monte Carlo simulations using the previously described 
EW distributions, but with $f$ now added as a free parameter.  We vary $f$ from 0 to 1 in steps of
0.01, and compute N=1000 simulations for each step.  We can then calculate the
probability distribution for $f$ given our N$_{obs}$=2 confirmed sources using Bayes' theorem:

\begin{equation}
p(f|N_{obs}=2) = \frac{p(N_{obs}=2|f) p(f)}{\int^1_0 p(N_{obs}=2|f) df}
\end{equation}

Here, $p(f)$ is the prior probability for $f$, which we take to be uniform for $0 \leq f \leq 1$, and $p(N_{obs}=2|f)$ is the probability,
drawn from our Monte Carlo simulations, that we would find N$_{obs}=2$ sources for a given value of $f$.  Assuming that
the intrinsic EW distribution for our observed sources is that of Paper II at $z=6$, we find $f =0.45 \pm 0.20$, 
while using the $z=7$ extrapolated distribution yields $f=0.34^{+0.24}_{-0.15}$.  In the Figure 5, we plot
the value of X(Ly$\alpha$) in the same luminosity bins of Paper II, as predicted by our best fit 
values of $f$.  

We stress that this figure is intended to serve as a continuation of the visualization provided in Papers I and II,
rather than a statistiscal result of our study.  Due to our strongly varying limiting EW sensitivity (as a function of 
both wavelength and object magnitude), choosing a fixed EW limit will exclude a non-negligible fraction of useful 
data from our analysis.  Our Monte Carlo simulations are able to utilize the full data set, simulating whether we would
have likely seen a line even when our EW limits are above the fixed thresholds used in Fig. 5, and thus represent the
major statistical result of this study.

Using the models of \cite{McQuinn07} to 
predict what global neutral hydrogen fraction, $\rm{X}_{HI}$ would be required to account for this decline, we find 
$\rm{X}_{HI}\simeq 0.44$, and $\rm{X}_{HI}\simeq 0.51$, respectively.  The models of \cite{Dijkstra11},
which provide a more comprehensive treatment of Ly$\alpha$ radiative transfer through outflows, result in
an increased value for $\rm{X}_{HI}$ in both cases.

\begin{figure}
\begin{center}
\includegraphics[width=0.5\textwidth]{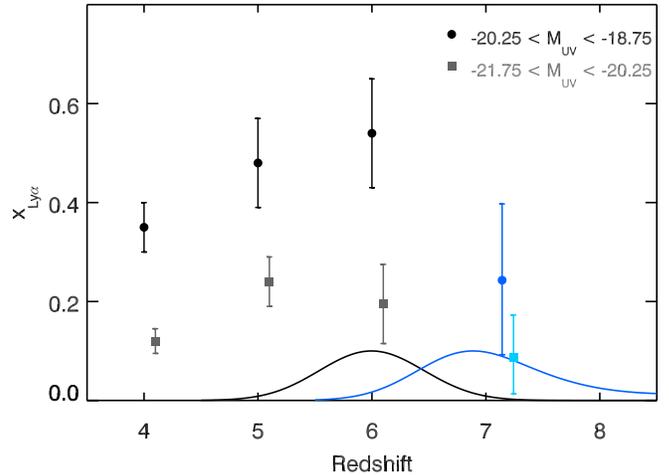}
\caption{The redshift-dependent fraction of color-selected Lyman break galaxies 
that reveal Ly$\alpha$ in emission, $X$(Ly$\alpha$), adjusted as discussed in the text to approximate
one within a similar luminosity range with a rest-frame EW in excess of 25 \AA.  Data points 
for the galaxies with $-21.75<\rm{M}_{UV}<-20.25$ are displaced by $+0.1$ in redshift for clarity.
Data over $4<z<6$ is from Paper I and Paper II,  and new estimates beyond $z>6.3$ are derived from the 
present paper, including sources discussed by \citet{Fontana10}.  The curves shown
represent the aggregate redshift probability distributions for our sources in the z $\simeq$ 6 bin (black), and
the z $\simeq$ 7 bin (blue); probability distributions for individual sources are typically much sharper.}
\label{fig:lya_frac}
\end{center}
\end{figure}

\vspace*{5mm}

\section{Discussion}

Although we consider the most likely explanation for our observed
decrease in the number of LBGs which show observable Ly$\alpha$
emission to be an increase with redshift in the neutral fraction of the IGM,
it is important to remember our assumptions. Foremost we have assumed
that {\it all} of our 26 targets have true redshifts beyond $z\simeq 6.3$. Should
there be low redshift interlopers or Galactic stars in our new sample, we will
overestimate the decline in the Ly$\alpha$ fraction. Secondly,
we have assumed the DEIMOS spectra from Paper II constitute a
representative sample for calculating the expected EW distribution
for 6.3 $<z<$ 8.2. Although the uncertainties here are not as great,
we plan further studies with DEIMOS to increase the statistical sample of 5.5 $<z<$ 6.3
LBGs.

Of course our observed decrease in the Ly$\alpha$ fraction could also
be attributed to an increased opacity arising from dust within the LBGs.
However, given the blue UV continuum slopes observed for galaxies
with $z>6.3$ \citep{Bouwens10a,Dunlop11}, we consider this explanation 
unlikely.

Our diagnosis of a possible increase in the neutral hydrogen fraction beyond $z\simeq$ 6.3 
is supported by the earlier study of \citet{Fontana10}. They
found 1 marginal candidate out of 7 targets whereas we find 2 robust and 1 marginal cases
out of our 19 targets spanning a larger luminosity and redshift range. Our conclusion is also
supported by LAEs studies at $z=5.7$ and $6.5$ by \cite{Ouchi10} and \cite{Kashikawa11}.  
Compared to $z=5.7$, their LAE sample at  $z=6.5$ displays systematically lower EWs for 
Ly$\alpha$.  They also derive little evolution in the rest UV luminosity function for LAEs, but a 
decrease in the Ly$\alpha$ LF,  which could be explained by an increase in X$_{HI}$.  Our derived
values of X$_{HI}$ are slightly higher than that of \cite{Kashikawa11}, perhaps consistent with our survey probing
to higher redshifts than their $z=6.5$ LAEs.  \cite{Hayes11} have recently compiled results 
from numerous Ly$\alpha$ and UV luminosity function studies to derive a volumetrically averaged Ly$\alpha$
escape fraction, and find very similar results.  Their derived Ly$\alpha$ escape fraction steadily increases with redshift below $z = 6$, then 
tentatively drops off at higher redshifts.

Very recently, \cite{Ono11} report the 
convincing detection of Ly$\alpha$ emission in a small fraction (3/11) of LBGs that, by virtue of their selection
using Subaru imaging, are more luminous ($\rm{M}_{UV}<-21$) than most of the objects considered here.  Such
a complementary campaign targeting luminous LBGs selected from larger volumes will provide further
insight into whether reionization is responsible for the declining fraction of line emission.

We note that our measured decrease in the fraction of LBGs with strong Ly$\alpha$ potentially
agrees with the result of \cite{Cowie11}.  Although they argue against any evidence for reionization
at $z = 6.5$, they find that  $\sim24 \%$ of galaxies at this redshift show strong Ly$\alpha$ emission,
comparable to the fraction we detect in this work, spread across a larger redshift range.

With the new generation of multi-object, near infrared spectrographs, such as MOSFIRE,
set to come online soon, the prospects for this field are bright.  In addition to the significant
multiplexing advantage, the increased sensitivity of these detectors will allow us to probe
the lower luminosity ranges at $z\geq 6.5$ to EW limits comparable to those in Paper II between
sky lines.  Having such a statistical sample is key for allowing the quantification of any change in
the hydrogen neutral fraction. 

\subsection*{ACKNOWLEDGMENTS}
We thank the referee for valuable comments which improved the manuscript, George Becker for allowing us to 
use his NIRSPEC reduction pipeline, and Mark Dijkstra for his helpful discussion.
RSE and MAS thank the Institute of Astronomy, Cambridge, where this work was
completed, for its support. We also wish
to recognize and acknowledge the very signiﬁcant cultural role and reverence that the summit of Mauna Kea
has always had within the indigenous Hawaiian community. We are most fortunate to have the opportunity to
conduct observations from this mountain.  DPS acknowledges financial support from a postdoctoral fellowship from
the Science Technology and Research Council. BER is supported by a Hubble Fellowship grant, program number HST-HF-51262.01-A provided 
by NASA from the Space Telescope Science Institute, which is operated by the Association of Universities for 
Research in Astronomy, Incorporated, under NASA contract NAS5-26555.  JSD acknowledges the support of the Royal Society via a Wolfson Research Merit award, and the support of the European Research Council via an Advanced Grant.  RJM acknowledges the support of the Royal Society via a University Research Fellowship.

\end{document}